\documentclass[reprint,superscriptaddress,amsmath,amssymb,prl]{revtex4-1} 

\usepackage{float}
\usepackage{url}
\usepackage{xcolor}
\usepackage{graphicx}
\usepackage{dcolumn}
\usepackage{bm}
\usepackage{changes} 
\usepackage[colorlinks=true, linkcolor=blue, citecolor=blue, urlcolor=blue]{hyperref}
\usepackage{units}
\newcommand{\HUB}{Institut f\"ur Physik \& IRIS Adlershof,
  Humboldt-Universit\"at zu Berlin, Newtonstra{\ss}e 15, 12489 Berlin, Germany}
\newcommand{\ZIB}{Zuse Institute Berlin, Takustra{\ss}e 7, 14195 Berlin, Germany}
\newcommand{\JCM}{JCMwave GmbH, Bolivarallee 22, 14050 Berlin, Germany}

\begin{document}

\title{Riesz-projection-based theory of light-matter interaction in dispersive nanoresonators}

\author{Lin Zschiedrich}
\affiliation{\JCM}
\author{Felix Binkowski}
\affiliation{\ZIB}
\author{Niko Nikolay}
\affiliation{\HUB}
\author{Oliver Benson}
\affiliation{\HUB}
\author{G\"unter Kewes}
\affiliation{\HUB}
\author{Sven Burger}
\affiliation{\JCM}
\affiliation{\ZIB}


\begin{abstract}
  We introduce a theory to analyze the
  behavior of light emitters in nanostructured environments rigorously. 
  Based on spectral theory, the approach opens the possibility
  to quantify precisely how an emitter decays to resonant states of the structure
  and how it couples to a background, also in the presence of general dispersive media.
  Quantification on this level is essential for designing and analyzing topical nanophotonic
  setups, e.g., in quantum technology applications.
  We use a numerical implementation of the theory for computing modal and background decay rates
  of a single-photon emitter in a diamond nanoresonator. 
\end{abstract}


\maketitle

{\it Introduction.}\quad
Resonance phenomena are omnipresent in physics. Storage and transfer of energy between different
resonant states allows one to explore wave effects in atomic, molecular, and optical physics as well as in
nuclear and condensed-matter physics and in other fields of science. 
Optical resonators are scaled down to  the wavelength scale and below by using modern nanotechnology,
as demonstrated in various material
systems~\cite{Vahala_OpticalMicrocav_2003, Bharadwaj_OptAntennas_2009}, including
plasmonic~\cite{Giannini_PlasmonNanoantennas_2011} and dielectric 
structures~\cite{Kuznetsov_DielectricNanostruc_2016}.
Placing pointlike sources in the vicinity of such nanoresonators or antennas enables 
exploration of new regimes of light-matter interaction.
Examples are single-photon emission with high
directivity~\cite{Curto2010UnidirectionalNanoantenna.,Ding_single_photons2016,Koenderink_SinglePhoton_2017},
nanoscopic plasmon lasers~\cite{Oulton2009,nogoniov2018nm}, and modification of chemical 
reaction rates by exploiting strong coupling in microcavities~\cite{Hutchison_Chem_2012}.

Theoretical models of light-matter interaction are needed to understand and optimize the performance of
related photonic devices.
Maxwell's equations can be solved directly to obtain solutions for the electromagnetic field.
For a deeper insight into physical properties, it is a common approach to use a modal description.
The resonant response of metallic nanostructures is governed by surface plasmon polaritons.
High-index dielectrics hosts electric and magnetic Mie-like modes which can be exploited in antenna
design~\cite{Kuznetsov_DielectricNanostruc_2016}.
For understanding the interaction of emitters with nanoresonators,
it is essential to precisely describe the coupling of the emitter to specific
modes~\cite{Hughes_PRA_2016,Feichtner_ModeMatching_2017}.
This coupling is quantified by individual modal Purcell 
factors~\cite{Purcell_1946,Sauvan_QNMexpansionPurcell_2013}.
Thus, in most approaches, the study of the (eigen-)modes and associated eigenfrequencies of the
resonating structure is essential.

The simplest model for modal analysis, a closed nondissipative system, yields a Hermitian
linear operator with a complete set of orthogonal eigenmodes.
By duality and based on a scalar product, these eigenmodes also serve as projectors which allow for an
expansion of the electromagnetic field into a sum of eigenmodes to characterize the light-matter interaction.
In the past decades, the more challenging study of open systems, which are usually described by non-Hermitian
operators, is an often addressed research topic in various applications including quantum
mechanics~\cite{Hislop_BookSpecTheoSchroed_1996, Zworski99resonancesin,bindel_zworski_toolboch_2006,
Calderon_ResonantStates_2010} and nano-optics \cite{Muljarov_EPL_2010,Sauvan_QNMexpansionPurcell_2013,
Kristensen_ModeVol_2014,Muljarov_PRB_2016,Lalanne_ReviewArxivQNM_2018}.
In a nutshell, the concept of eigenmodes has been generalized to the theory of resonant states,
also called quasinormal modes (QNMs).
QNMs are orthogonal with respect to an unconjugated scalar product~\cite{Leung_CompleteOrthQNM_1994}
which allows identification of QNMs with projectors again. Also, in the case of dispersive
materials which are ubiquitously present in nano-optical 
resonators~\cite{Giannini_PlasmonNanoantennas_2011, Hughes_Optica_2017},
there exist approaches for QNM expansion \cite{Sauvan_QNMexpansionPurcell_2013,Muljarov_PRB_2016}.
However, the orthogonality and normalization of the QNMs, especially in the case of dispersive media, are
still under active research and discussed controversially in the 
literature~\cite{Sauvan_QNMexpansionPurcell_2013,
Kristensen_NormQNM_2015,Sauvan_procSPIE_2015,Muljarov_Purcell_PRB_2016,
Muljarov_PRB_2016,Muljarov_PRA_Comment_2017}.
The discrete set of QNMs is supplemented by the continuous spectrum of the operator capturing the
nonresonant background scattering~\cite{PhysRevA.89.023829}.
State-of-the-art approaches using QNM expansion do not incorporate the continuous spectrum.
These can well be applied when coupling to the background is
negligible~\cite{Sauvan_QNMexpansionPurcell_2013}.
However, important application classes rely on designs with significant background coupling
which is present when low-quality ($Q$) factor resonances are involved~\cite{Purcell_1946}.
For realizing integrated single-photon sources, the involved resonant states are preferably at
low $Q$ factor, enabling fast, pulsed operation~\cite{Gschrey_QDot_2015}.
Also, for modifying photochemical reactions, coupling of molecules to resonant states with a low $Q$ factor
is used due to better accessibility compared to high $Q$ factor resonances~\cite{Hutchison_Chem_2012}.
Theoretical description and numerical optimization of related setups therefore
essentially require precise treatment and precise distinction of
coupling to the background and to the resonant states.

Riesz projections (RPs) can be used to compute these quantities in an elegant way.
RPs are a well-known concept in spectral theory~\cite{Hislop_BookSpecTheoSchroed_1996} 
and they do not rely on orthogonality relations and the explicit knowledge of eigenfunctions.
RPs are based on contour integration and provide a powerful means to analyze the spectrum of
partial differential operators. Note that parallel to this work, a scalar product involving
auxiliary fields has been proposed to ensure the orthogonality of QNMs for typical dispersive 
media~\cite{Yan_PRB_2018}.

In this work, we present a theory for modeling dispersive light-matter interaction based on RPs.
We show that RPs can also be used to model
the nonresonant background interaction in a closed form.
The theory allows for a straightforward numerical implementation which essentially requires solving
\mbox{time-harmonic} scattering problems for complex frequencies. We apply the method to compute modal
decay rates of a dipole emitter embedded in a diamond nanodisk antenna showing a weak coupling to the
QNMs and a significant background coupling.

\begin{figure}
	\centering
	{\includegraphics[width=0.45\textwidth]{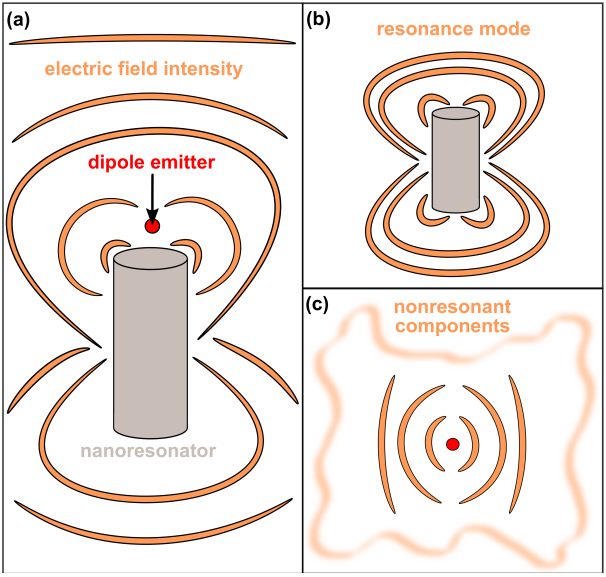}}\\
	\caption{Schematic decomposition of the electromagnetic field caused by a dipole emitter
		in the vicinity of a nanoresonator.
		(a) Total electromagnetic field.
		(b) A resonance mode of the nanoresonator. (c) Nonresonant components of the electromagnetic field.
		This part includes also the singularity resulting from the dipole source.}
	\label{fig:nanodisc}
\end{figure}

\begin{figure}
	\centering
	{\includegraphics[width=0.45\textwidth]{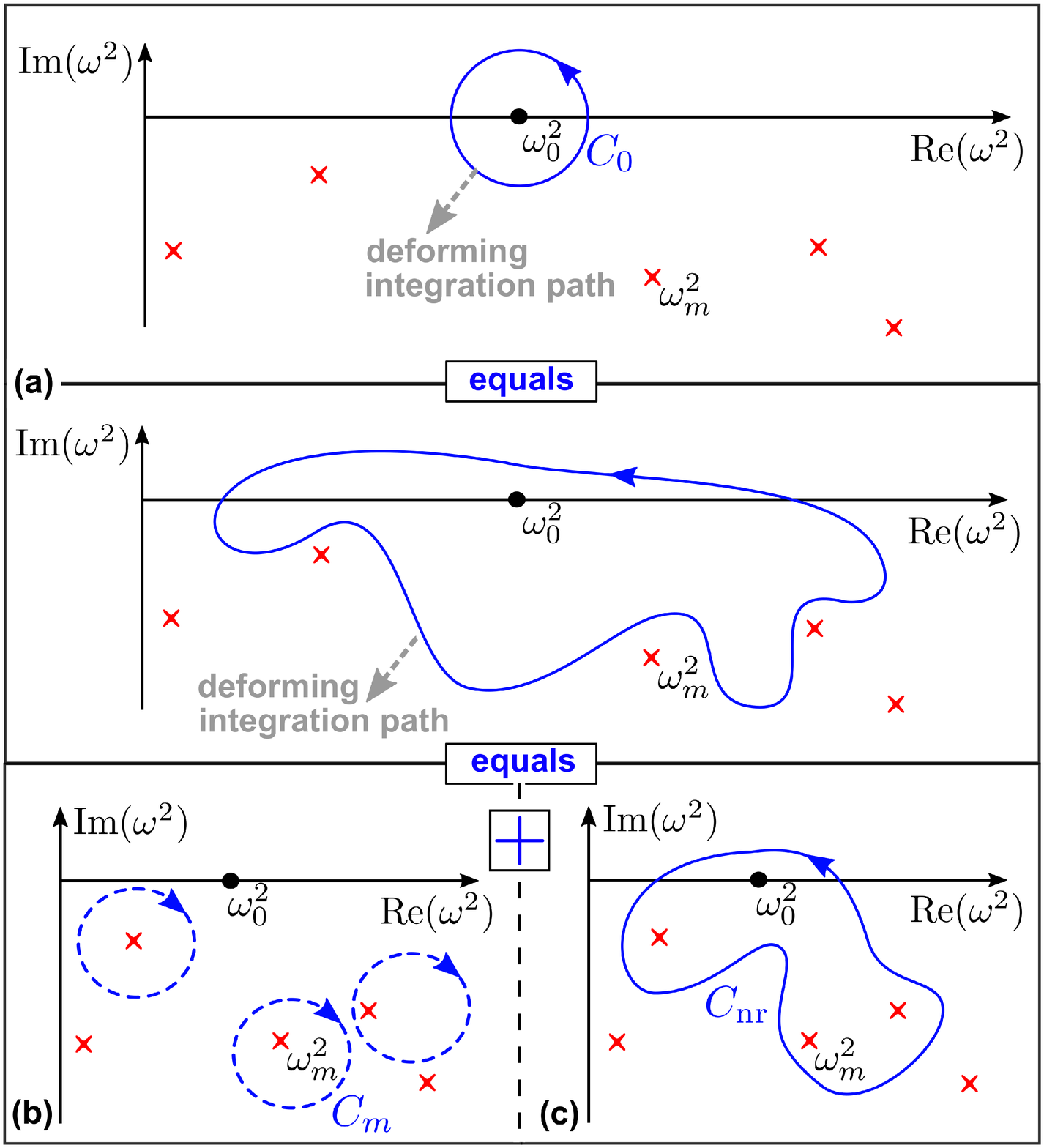}}
	\caption{Contour integration in the complex $\omega^2$ plane for computing the Riesz projection expansion,
		Eq.~\eqref{expansion_complete}.
		The red crosses represent resonance poles ${\omega}^2_m$, the blue curves are the
		integration curves for Eqs$.$ (\ref{cauchy})-(\ref{nonres_comp}).
		(a)~Top: Integration path $C_0$ around $\omega_0^2$, see Eq.~\eqref{cauchy}.
		Bottom: Deforming the integration path without enclosing resonance poles does not modify the integral.
		(b)~Integration curves ${C}_{m}$ in negative direction for computing Riesz projections,
		see Eq.~\eqref{riesz_proj}.
		(c)~Outer integration path $C_\textrm{nr}$ for quantifying the interaction with nonresonant
		components, see Eq.~\eqref{nonres_comp}.}
	\label{fig:cauchy}
\end{figure}

{\it Riesz projection expansion.}\quad
In the following, we consider electromagnetic fields in the vicinity of optical nanostructures,
as illustrated in Fig.~\ref{fig:nanodisc}.
The total field, sketched in Fig.~\ref{fig:nanodisc}\hyperref[fig:nanodisc]{(a)},
is decomposed into resonant and nonresonant components
[see Figs.~\ref{fig:nanodisc}\hyperref[fig:nanodisc]{(b)} and~\hyperref[fig:nanodisc]{1(c)}].
In the steady-state regime, the corresponding electric fields $\mathbf{E}(\mathbf{r},\omega)$
are solutions to the time-harmonic Maxwell's equations in the second-order form
\begin{align}
  \mathbf{\nabla} \times \mu^{-1}  \mathbf{\nabla} \times \mathbf{E}(\mathbf{r},\omega) -
  \omega^2\epsilon(\omega) \mathbf{E}(\mathbf{r},\omega) =
    i\omega\mathbf{J}(\mathbf{r}), \label{maxwell_eq}
\end{align}
where $\omega \in \mathbb{C}$ is a complex angular frequency.
The material dispersion is described by the permittivity tensor $\epsilon(\omega)$,
and the permeability tensor $\mu$ typically equals the vacuum permeability $\mu_0$.
The source term $\mathbf{J}(\mathbf{r})$ relates 
to impressed currents. For open problems, Eq.~\eqref{maxwell_eq} is equipped with outgoing
radiation conditions which can be realized by complex scaling in space of the corresponding partial differential
operator. Incident exterior light sources can be incorporated in $\mathbf{J}(\mathbf{r})$. 
Physically relevant scattering solutions have real frequencies $\omega_0\in \mathbb{R}$.
The fields $\mathbf{E}(\mathbf{r},\omega)$ can be regarded as an analytical continuation of
$\mathbf{E}(\mathbf{r},\omega_0)$ into the complex plane.
In this context, the QNMs correspond to complex frequencies ${\omega}_{m} \in \mathbb{C}$, $m=1,\dots,M$,
where $\mathbf{E}(\mathbf{r},\omega)$ has a resonance pole, i.e., a singularity.

To decompose $\mathbf{E}(\mathbf{r},\omega_0)$ into its resonant and nonresonant parts,
we consider the \mbox{$z = \omega^2$ plane} and write
\mbox{$\mathbf{E}(\mathbf{r},z) = \mathbf{E}(\mathbf{r},\omega = \sqrt{z})$}.
Cauchy's residue theorem gives
\begin{align}
  \mathbf{E}(\mathbf{r},\omega_0) =
  \frac{1}{2 \pi i} \oint \limits_{C_0} \frac{\mathbf{E}(\mathbf{r},z)}{z-\omega_0^2} \text{ d}z,
  \label{cauchy}
\end{align}
where $C_0$ is a closed curve around $\omega_0^2$ so that $\mathbf{E}(\mathbf{r},z)$ 
is holomorphic inside of $C_0$, as shown in Fig.~\ref{fig:cauchy}\hyperref[fig:cauchy]{(a)}.
Then, deforming the path of integration so that an outer curve $C_{\text{nr}}$
includes $\omega_0^2$, the resonance poles \mbox{${\omega}_1^2,\dots,{\omega}_M^2$}
 and no further poles
yields
\begin{align}
\oint\limits_{C_0} \frac{\mathbf{E}(\mathbf{r},z)}{z-\omega_0^2} \text{ d}z
=  &- \oint\limits_{{C}_1} \frac{\mathbf{E}(\mathbf{r},z)}{z-\omega_0^2} \text{ d}z -
\dots - \oint\limits_{{C}_M} \frac{\mathbf{E}(\mathbf{r},z)}{z-\omega_0^2} \text{ d}z \nonumber \\
&+	\oint\limits_{C_{\text{nr}}} \frac{\mathbf{E}(\mathbf{r},z)}{z-\omega_0^2} \text{ d}z, \nonumber
\end{align}
see Figs$.$~\ref{fig:cauchy}\hyperref[fig:cauchy]{(b)} and~\hyperref[fig:cauchy]{2(c)}.
Thereby, we obtain the expansion
\begin{align}
  \mathbf{E}(\mathbf{r},\omega_0) = \sum_{m=1}^{M} {\mathbf{E}}_m (\mathbf{r},\omega_0) +
         {\mathbf{E}}_{\text{nr}} (\mathbf{r},\omega_0),
         \label{expansion_complete}
\end{align}
where the fields
\begin{align}
  {\mathbf{E}}_m (\mathbf{r},\omega_0) = -\frac{1}{2 \pi i}
  \oint\limits_{{C}_m} \frac{\mathbf{E}(\mathbf{r},z)}{z-\omega_0^2} \text{ d}z \label{riesz_proj}
\end{align}
are related to the resonance poles \mbox{${\omega}_1^2,\dots,{\omega}_M^2$}.
The field
\begin{align}
  {\mathbf{E}}_{\text{nr}} (\mathbf{r},\omega_0) =  \frac{1}{2 \pi i}
  \oint\limits_{C_{\text{nr}}}
  \frac{\mathbf{E}(\mathbf{r},z)}{z-\omega_0^2} \text{ d}z
  \label{nonres_comp}
\end{align}
quantifies the nonresonant components and contributions from possible resonance poles outside of
the integration curve $C_{\text{nr}}$. It has to be ensured that $C_\mathrm{nr}$ does not
cross the branch cut in the $z = \omega^2$ plane starting from $z=0$.
The fields in Eq.~\eqref{riesz_proj} are essentially
RPs applied to Eq.~\eqref{maxwell_eq}; see results from spectral theory~\cite{Hislop_BookSpecTheoSchroed_1996}.
The RP expansion offers a general physical understanding of resonance phenomena without the need to
normalize exponentially diverging fields. 
Clearly, the integrals in Eqs.~\eqref{riesz_proj} and~\eqref{nonres_comp} are independent of the
particular choice of the contours ${C}_m$ and $C_{\text{nr}}$.
Therefore, precise locations of the resonance poles are not required.
Also, when a contour includes multiple resonance poles the contour integral gives the projector
onto the space of corresponding QNMs.
In this way, it is possible to construct projectors for frequency ranges without
detailed \mbox{\textit{a priori}} knowledge. 
This case implies that a specific choice of the number $M$ in 
Eq.~(\ref{expansion_complete}) is not necessary.

RP expansion can be applied to any light source; however, of special interest are pointlike sources.
These can be modeled as dipole 
emitters \mbox{$\mathbf{J}(\mathbf{r}_0) = \mathbf{j}\delta(\mathbf{r}-\mathbf{r}_0)$},
where $\mathbf{j} = -i \omega \mathbf{p}$ with dipole moment $\mathbf{p}$ at position $\textbf{r}_0$.
Its enhanced emission rate in the vicinity of a nanoresonator is characterized by
the Purcell factor~\cite{Purcell_1946}, also termed normalized decay rate,
\begin{align}
  {\Gamma}(\omega_0) = -\frac{1}{2} \text{Re}\left({\textbf{E}}(\textbf{r}_0,\omega_0) \cdot
  \mathbf{j}^*\right)/\Gamma_\text{b}, \label{Purcell}
\end{align}
where $\Gamma_\text{b}$ is the decay rate of the emitter in homogeneous background 
material~\cite{Sauvan_QNMexpansionPurcell_2013,ZschiedrichGreinerBurgeretal.2013}.
To quantify the coupling of the emitter to each of the single RPs
${\textbf{E}}_m(\textbf{r},\omega_0)$, we introduce the modal normalized decay rate
\begin{align}
  {\Gamma}_m(\omega_0) = -\frac{1}{2} \text{Re}\left({\textbf{E}}_m(\textbf{r}_0,\omega_0) \cdot
  \mathbf{j}^*\right)/\Gamma_\text{b}.
  \label{Purcell_modal}
\end{align}
The nonresonant normalized decay rate is analogously given by
\begin{align}
  {\Gamma}_{\text{nr}}(\omega_0) = -\frac{1}{2} \text{Re}\left({\textbf{E}}_{\text{nr}}(\textbf{r}_0,\omega_0)
  \cdot \mathbf{j}^*\right)/\Gamma_\text{b}.
  \label{Purcell_nr}
\end{align}

\begin{figure}[]
	\centering
	{\includegraphics[width=0.5\textwidth]{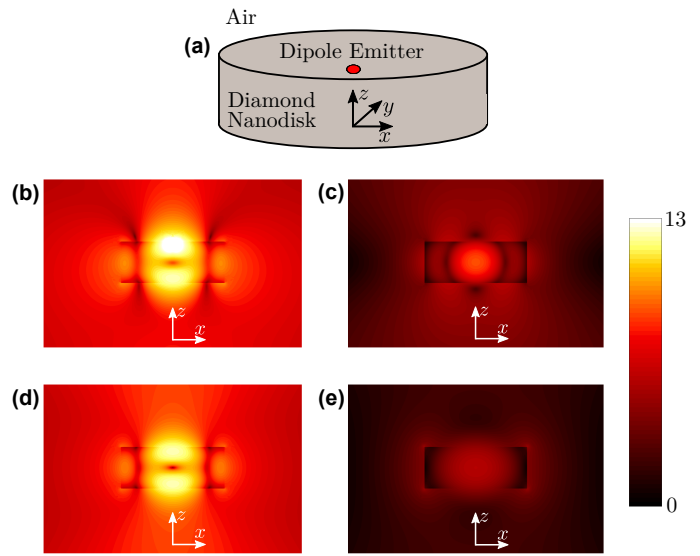}}
	\caption{(a)~Sketch of a diamond nanodisk antenna (diameter $\unit[400]{nm}$, height $\unit[160]{nm}$) in air
		with an embedded dipole emitter placed $\unit[15]{nm}$ below the upper surface
		and on the cylindrical symmetry axis ($z$ axis).
		The dipole with normalized strength is oriented in \mbox{$x$ direction} and oscillates at frequency
		\mbox{$\omega_0 = 2\pi c/\unit[455]{nm}$}.
		(b)~Log~plot (a.u.)~of the total electric field intensity.
		\mbox{(c)-(e)}~Log~plots (a.u.)~of the electric field intensity of
		the three RPs for the complex eigenfrequencies $\omega_1 = 2\pi c/(\unit[406+16i]{nm})$,
		$\omega_2=2\pi c/(\unit[454+13i]{nm})$, and $\omega_3=2\pi c/(\unit[655+55i]{nm})$, respectively.}
	\label{fig:riesz_cart}
\end{figure}

{\it Numerics.}\quad
For the numerical realization of the RP expansion, we calculate the contour integrals in
Eq.~(\ref{riesz_proj}) and~(\ref{nonres_comp}) using a simple trapezoidal rule.
At each integration point, it is required to solve Eq.~(\ref{maxwell_eq}) for a complex frequency which is
done with a finite-element method (FEM solver, \mbox{JCMsuite}). Perfectly matched layers
(PMLs)~\cite{Berenger:1994:PML:1718379.1718399} are used to enforce outgoing radiation conditions.
To ensure an accurate FEM discretization for singular dipole sources,
we use a subtraction field approach
\mbox{$\mathbf{E}(\mathbf{r},\omega) = \mathbf{E}_\mathrm{b}(\mathbf{r},\omega) +
  \mathbf{E}_\mathrm{c}(\mathbf{r},\omega)$}.
The field $\mathbf{E}_\mathrm{b}(\mathbf{r},\omega)$ is the analytically given solution to a 
dipole source in homogeneous bulk material.
The correction field
$\mathbf{E}_\mathrm{c}(\mathbf{r},\omega)$ is 
suitable for an accurate FEM discretization \cite{ZschiedrichGreinerBurgeretal.2013}.
Furthermore, for problem set\-ups with geometries of cylindrical symmetry, we reduce the three-dimensional computation to a
series of two-dimensional simulations with angular modes $e^{i n_\varphi \varphi}$.
The resonance poles are computed with a linear eigenvalue solver using an augmented field formulation.
However, as mentioned above, the precise locations of the eigenvalues could be replaced by rough guesses.
Self-adaptive approaches can be used for constructing suitable integration paths and to avoid crossing
resonance poles. To run scans of the frequency $\omega_0$
in the range $\left[\omega_{\text{min}},\dots, \omega_{\text{max}}\right]$,
note that the integrand in the RPs,  Eq.~\eqref{riesz_proj}, only depends on $\omega_0$ 
by the factor $1/(z-\omega_0^2)$.
Therefore, for the whole scan,  the fields $\mathbf{E}(\mathbf{r},\omega)$ need to be evaluated
only once at each integration point.
Furthermore, all calculations can be performed in parallel.
Due to these properties, the numerical realization is remarkably fast.
We mention that RPs have also been used for algebraic eigenvalue solvers~\cite{Asakura_JSIAMlett_2009,
Beyn_LinAlAppl_2012,Gavin_JcompPhys_2018}.

\begin{figure}[]
	\centering
	{\includegraphics[width=0.5\textwidth]{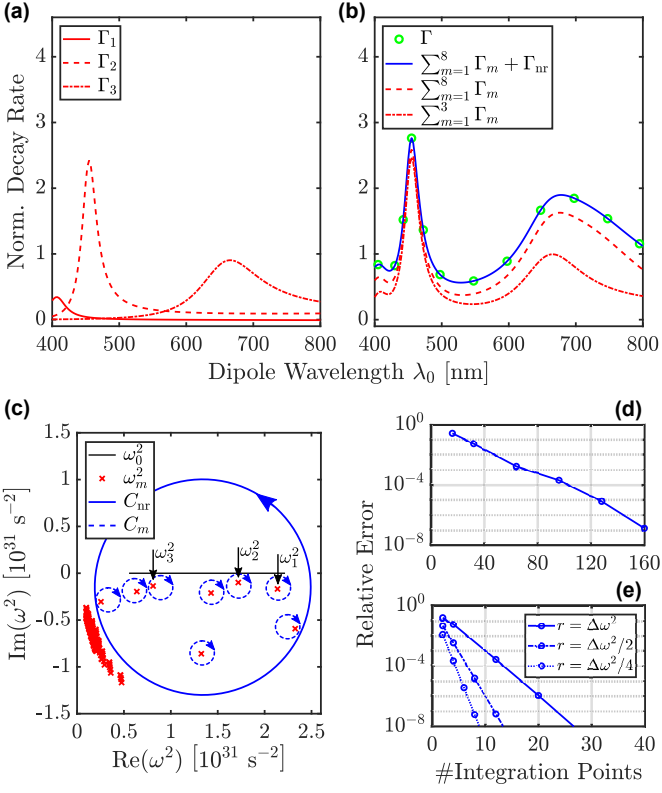}}
	\caption{Numerical results for the nanoantenna shown in
		Fig.~\ref{fig:riesz_cart}\hyperref[fig:riesz_cart]{(a)}.
		(a)~\mbox{${\Gamma}_m$ spectra} of the three dominant RPs from
		Figs.~\ref{fig:riesz_cart}\hyperref[fig:riesz_cart]{(c)}-\hyperref[fig:riesz_cart]{3(e)}.
		(b)~Normalized decay rates: $\Gamma$ (quasiexact solution),
		$\sum_{m}\Gamma_m + \Gamma_\mathrm{nr}$ (complete RP expansion),
		$\sum_{m=1}^M\Gamma_m$ (RPs for first $M$ resonance poles).
		(c)~Resonance poles of the nanoresonator and integration curves
		for computing $\Gamma_m$ ($C_m$ not to scale) and $\Gamma_\mathrm{nr}$.
		(d)~Convergence of $\Gamma_{\text{nr}}$ with respect to $C_\textrm{nr}$: 
		Maximum relative error $\left(\sum_{m=1}^{8}\Gamma_m + \Gamma_\mathrm{nr}-\Gamma\right)/\Gamma$
		as a function of the number of integration points [integration path as in (c)].  
		(e)~Convergence of \mbox{$\Gamma_2(\omega_0 = 2\pi c/\unit[455]{nm})$}
		with respect to the numerical parameters of the contour integration: 
		Relative error of $\Gamma_2$ as a function of the number of integration points, for different
		integration paths (circle $C_2$ with radii $r$ where 
		$\Delta \omega^2 = 5\times 10^{29} \text{s}^{-2}$),
		reference solution computed with $r= \Delta \omega^2/4$, and 64 integration points.}
	\label{fig:numerics}
\end{figure}

{\it Application.}\quad
Next, we apply the presented method to a highly topical example: a stable solid-state
emitter in a nanoantenna.
There is an urgent need for such systems to be used as single-photon sources for optical
quantum technologies~\cite{andersen2018acs}.
Room-temperature operation and directional emission at high rate are mandatory.
As diamond is known to host various interesting defects; we consider a setup where a
nitrogen-vacancy (N-V) center~\cite{SIMO:SIMO255} is hosted in a (dielectric)
diamond nanoantenna.
An all-diamond realization would be ultracompact and ideal for large-scale integration.
The specific geometry is depicted in Fig.~\ref{fig:riesz_cart}\hyperref[fig:riesz_cart]{(a)}.
The dipole emitter is placed on the symmetry axis and polarized in the $xy$ plane;
therefore only angular modes with $n_\varphi = \pm 1$ are populated.
The diamond permittivity $\epsilon(\omega)$ is described by a two-pole Lorentz model
\mbox{$\epsilon(\omega) =  \epsilon_0(1+\epsilon_{p_1}+\epsilon_{p_2})$},
where
\mbox{$\epsilon_{p_{1,2}}=\Delta_{\epsilon_{1,2}}\omega_{p_{1,2}}^2/(\omega_{p_{1,2}}^2-2i\omega\gamma-\omega^2)$},
with
$\Delta_{\epsilon_{1}} = 0.3306$,
$\Delta_{\epsilon_{2}} = 4.3356$, 
$\omega_{p_{1}} = 2\pi c/\unit[175]{nm}$,
$\omega_{p_{2}} = 2\pi c/\unit[106]{nm}$,
and damping \mbox{$\gamma = 0$}~\cite{Peter1923}.

We investigate the device within the wavelength range of
\mbox{$\lambda_0 \in \left[\unit[400]{nm},\dots,\unit[800]{nm}\right]$}.
For computing the RPs, we use four integration points in Eq.~\eqref{riesz_proj}.
Figure~\ref{fig:numerics}\hyperref[fig:numerics]{(a)} shows the modal normalized decay
rates $\Gamma_m$ of the three RPs corresponding to the resonance poles with smallest imaginary parts.
Each spectrum shows a maximum at the wavelength corresponding to the real part of the respective pole.
The highest decay rate is observed at around \mbox{$\lambda_0 = \unit[455]{nm}$}.
The RPs at $\omega_0=2\pi c/\lambda_0$ for three significant eigenfrequencies $\omega_m$
are shown in Figs.~\ref{fig:riesz_cart}\hyperref[fig:riesz_cart]{(c)}-\hyperref[fig:riesz_cart]{3(e)}.
For comparison, the total field solution computed from Eq.~(\ref{maxwell_eq}) is shown in
Fig.~\ref{fig:riesz_cart}\hyperref[fig:riesz_cart]{(b)}.
We note that for the investigated case where a single pole is enclosed in each contour integral,
the RP is a multiple of the corresponding QNM.
However, as mentioned above, from the
QNMs only it is not possible to compute modal expansion coefficients without an orthogonality relation,
i.e., without scalar products, yielding a
separation of the Maxwell's equations in a modal sense.
The presented approach is not restricted to specific geometrical setups. 
  Therefore, also handling complex environments of the nanodisk antenna, including, e.g., 
  layered structures, waveguides, and arbitrarily shaped objects, is straightforward.

Figure~\ref{fig:numerics}\hyperref[fig:numerics]{(b)} validates the completeness of the expansion
in Eq.~(\ref{expansion_complete}).
The quasiexact solution $\Gamma$ is gained from solving the scattering problem in Eq.~(\ref{maxwell_eq})
and applying Eq.~(\ref{Purcell}).
Using Eq.~\eqref{riesz_proj} and Eq.~\eqref{Purcell_modal} for the first three resonance poles yields 
an incomplete RP expansion $\sum_{m=1}^{3}\Gamma_m$
which already reproduces the characteristics of $\Gamma$.
Using the first eight poles, the agreement of the incomplete RP expansion
$\sum_{m=1}^{8}\Gamma_m$
with the quasiexact solution improves.
Adding the nonresonant part $\Gamma_\mathrm{nr}$,
calculated with Eq.~\eqref{Purcell_nr},
gives the theoretically expected match to the
quasiexact solution.
Here, for the computation of  $\mathbf{E}_\mathrm{nr}(\mathbf{r}, \omega_0)$ in Eq.~\eqref{nonres_comp},
we use 128 integration points.
We attribute the fact that the nonresonant components are of significant quantitative impact to the
nature of the diamond nanodisk antenna.
Due to its relatively low refractive index,
the structure hosts many weakly localized modes of low $Q$ factor.
Thus, the
coupling to the background continuum of modes plays an important
role for the Purcell factor.

Figure~\ref{fig:numerics}\hyperref[fig:numerics]{(c)} details the position of the resonance
poles in the complex plane and the used contour integral curves for this example.
We distinguish between physical resonance poles and so-called
PML poles~\cite{PhysRevA.89.023829,Yan_PRB_2018}.
Physical poles are stable with respect to a change of the numerical parameters and are
therefore related to the discrete part of the operator spectrum.
The PML poles stem from the continuous part of the spectrum of the operator and yield
algebraic eigenvalues due to the discretization and truncation of the open resonator system.
In this sense, the integral over the outer contour $C_\mathrm{nr}$ comprises the continuous part
of the operator (PML modes) as well as further QNMs which might be present outside of $C_\mathrm{nr}$.
Note that the bulk emission term $\mathbf{E}_\mathrm{b}(\mathbf{r},\omega)$ in the subtraction
field approach \mbox{$\mathbf{E}(\mathbf{r},\omega) = \mathbf{E}_\mathrm{b}(\mathbf{r},\omega)
  + \mathbf{E}_\mathrm{c}(\mathbf{r},\omega)$} is an analytic function in the entire complex plane.
Hence, $\mathbf{E}_\mathrm{b}(\mathbf{r},\omega)$ does not contribute to the RPs
$\mathbf{E}_m(\mathbf{r},\omega_0)$, which are therefore smooth fields, cf., 
Figs.~\ref{fig:riesz_cart}\hyperref[fig:riesz_cart]{(c)}-\hyperref[fig:riesz_cart]{3(e)},
whereas $\mathbf{E}_\mathrm{nr}(\mathbf{r},\omega_0)$ produces the singularity. 

The numerical efficiency of the RP expansion depends on the numerical convergence of the contour
integral with respect to the number of integration points.
For the outer contour $C_\mathrm{nr}$, as plotted in
Fig.~\ref{fig:numerics}\hyperref[fig:riesz_cart]{(c)}, we observe convergence with
respect to the number of integration points, see Fig.~\ref{fig:numerics}\hyperref[fig:riesz_cart]{(d)}.
For the contours of the single RPs,
we verified that four integration points are sufficient to reach a relative accuracy
of the derived modal decay rate better than $10^{-6}$,
see Fig.~\ref{fig:numerics}\hyperref[fig:riesz_cart]{(e)}.

{\it Conclusions.}\quad
In conclusion, we presented a theoretical approach to explain the coupling of light sources to
dispersive nanoresonators by means of an electromagnetic field expansion with Riesz projections.
The method allows for the precise definition and computation of the field expansion into modal and
background parts and for the evaluation of linear functionals, e$.$g$.$, modal and background decay rates.
We applied the approach to model the coupling of an emitting defect center in diamond to a nanodisk antenna 
supporting several weakly localized resonant states.
The method is applicable to systems with any material dispersion obeying \mbox{Kramers-Kronig} relation.
We therefore expect that the approach will prove especially useful for understanding and
designing novel photonic devices
with material properties that can only be accurately modeled using
high-order rational fits to measured data. 
Riesz projection expansion further establishes a route for
quantitative modal analysis of omnipresent nano-optical systems
with relevant nonresonant background. 
  The presented concepts may also be applied to explore
  open resonators in other fields of physics, e.g., in
  phononic structures~\cite{hussein2014dynamics} and acoustic metamaterials~\cite{Cummer_acoustics_2016}.

We acknowledge support by the Einstein Foundation Berlin (ECMath, Project No.~OT9),
the Senat von Berlin (IBB 10160385, FI-SEQUR, cofinanced by the European EFRE program),
the German Research Foundation DFG (SFB\,951, Project No.~B2),
and the Ministry of Science and Education BMBF (Project No.~13N14148, Nano-Film).\\

\end{document}